\documentclass[12pt,a4paper]{article}
\usepackage[cp1251]{inputenc}
\usepackage[english, russian]{babel}
\usepackage{amssymb,amsmath}
\usepackage{graphicx}
\usepackage{ucs}

\begin{document}

\begin{center}
{ \bf \large The Ranking Problem of Alternatives as a Cooperative
Game}

\medskip
A.Yu. Kondratev and V.V. Mazalov

Institute of Applied Mathematical Research, Karelian Research
Center,
Russian Academy of Sciences\\
185910, Russia, Petrozavodsk, ul. Pushkinskaya, 11\\
\normalsize e-mail: kondratev@krc.karelia.ru, vmazalov@krc.karelia.ru \\

\end{center}

\bigskip

{\bf Abstract}. This paper considers the ranking problem of
candidates for a certain position based on ballot papers filled by
voters. We suggest a ranking procedure of alternatives using
cooperative game theory methods. For this, it is necessary to
construct a characteristic function via the filled ballot paper
profile of voters. The Shapley value serves as the ranking method.
The winner is the candidate having the maximum Shapley value. And
finally, we explore the properties of the designed ranking
procedure.

\bigskip


\bigskip

{\large \bf 1. Introduction}
\medskip

The present paper focuses on the ranking problem of candidates for
a certain position based on ballot papers filled by their supporters.
Such problem arises during elections of a president, a company's CEO,
a professor in a chair and many other positions. By an assumption,
elections are free, honest and open. In the course of voting,
electors fill ballot papers and specify their preferences for existing candidates.
Generally, the number of ballot papers appreciably exceeds the number of candidates.
The winner is defined on the basis of all filled ballot papers. Here a major role
belongs to the winner selection method. This ballot paper handling technique
must possess a series of positive properties. We will believe that elections
choose an appropriate candidate from at least two ones. Voters have to fill
a ballot paper and specify their relative preferences for given candidates.
As a matter of fact, there exist different ways of ballot paper filling.

In the case of {\bf majority voting}, electors have to specify a most preferable candidate.
The winner is the candidate receiving the simple majority of votes. This voting rule
appears widespread, has easy numerical implementation and requires reasonable computations.
However, it neglects voting aspects in situations when several candidates
are equally preferable for electors.

Under {\bf approval voting}, an elector specifies only trusted
candidates (see Brams, Fishburn 1978, 2005). Therefore, each voter
assigns $1$ to trusted candidates and $0$ to others. The candidate
with the maximum number of labels $1$ wins the elections. This
voting rule demonstrates higher complexity and insensitivity to the
existing preferences of electors. The idea with three labels $0, 1,
2$ was pioneered in the paper (Hillinger 2005). An axiomatization of
such score voting procedures (evaluating voting rules) was suggested
in (Gaertner, Xu 2012).

Moreover, there exist {\bf range voting rules}, where each candidate
receives an evaluation from 0 to 100 (W.D.Smith 2000). The numerical
implementation of this procedure causes no difficulties, either; but
it is relatively insensitive to the preferences for candidates.

According to {\bf the majority judgement approach}, each candidate
is associated with some group of preferences, e.g., $A,B,C,D,…$
(Balinski, Laraki 2007). Such method depends on the number of groups
and entails difficulties in the course of ballot paper handling, but
seems computationally easy.

{\bf Ranked elections} dictate placing all candidates in a ballot paper in the descending order
of voter's preferences. This method reflects the existing preferences of all voters for candidates
in the most accurate way. Several techniques to count the votes are applicable here,
and they have different sensitivity to possible variations in voters' preferences.

In this work, we examine ranking procedures. Consider given sets of
candidates $\{a,b,c,…\}$ and voters $\{1,2,…,n\}$. Each voter fills
a ballot paper by ranking all candidates in the descending order of
its preferences. It is required to define the winner based on the
filled ballot papers. There exist a series of procedures for ballot
papers handling and winner choice; in this context, we mention the
majority rule, the procedures suggested by Borda and Copeland (see
Klamler 2006), as well as the sequential pairwise comparison
procedure and dictatorship voting. In certain cases, it is necessary
to define several winners, e.g., to choose a management committee or
a team for a project. Here a possible approach is to address the
minimax and minisum procedures (Brams, et al. 2007, Kilgour 2010).

The present paper suggests involving some methods of cooperative game theory as
ballot paper handling procedures. The idea is to construct a special-form
characteristic function of such game using filled ballot papers,
with subsequent weighting of each player (e.g., by the Shapley value).
The stated procedure enjoys remarkable properties, as it takes into account
not just the correlation of two candidates, but the correlation of any candidates
depending on their belonging to certain coalitions. The procedure yields weights
for all candidates and is therefore applicable to models, where one has to choose
a group of winning candidates (instead of a single winner).

\bigskip

{\large \bf 2. The ranking procedure}
\medskip

Consider the voting problem, where $n \geq 2$ electors have to choose a winner
among $m\geq 2$ candidates. The preferences of each voter are defined by a linear order
on the set of candidates. Denote by $A=\{ a, b, c, \ldots \}$ and $P=\{ 1, 2, \ldots, n \}$
the sets of candidates and voters, respectively. Based on joint preferences, a voting procedure
(comprising a ballot paper and a counting rule of votes) leads to a result of voting.
A voting procedure aims at assigning ranks from 1 to $m$ to all candidates;
note that such ranking can be nonstrict.

Below we give an example of a voting procedure with 45 electors and 5 candidates.

{\bf Example~1.} There are $n=45$ voters and $m=5$ candidates.
The existing preferences of voters are defined by the following table.
\begin{center}
{\small\textit{Preference profile} \\
\vspace{3mm}
\begin{tabular}{|c|c|c|c|c|c|c|c|}
\hline
5 & 5 & 8 & 3 & 7 & 2 & 7 & 8 \\
\hline
a & a & b & c & c & c & d & e \\
c & d & e & a & a & b & c & b \\
b & e & d & b & e & a & e & a \\
e & c & a & e & b & d & b & d \\
d & b & c & d & d & e & a & c \\
\hline
\end{tabular}
}
\end{center}

Designate by $h(i,j)$ the number of ballot papers, where candidate $i$
is preferable to candidate $j$. Let us compile a matrix from the values
of the function $h(i,j)$. Such matrix is called {\bf a tournament matrix}.
For instance, the tournament matrix in Example~1 takes the following form.

\begin{center}
{\small\textit{Tournament matrix} \\
\vspace{3mm}
\begin{tabular}{|c|c|c|c|c|c|c|c|}
\hline
  & a & b & c & d & e  \\
\hline
a &   & 20 & 26 & 30 & 22  \\
\hline
b & 25 &   & 16 & 33 & 18  \\
\hline
c & 19 & 29 &  & 17 & 24  \\
\hline
d & 15 & 12 & 28 &   & 14  \\
\hline
e & 23 & 27 & 21 & 31 &    \\
\hline
\end{tabular}
}
\end{center}

Generally, tournament matrices serve for the final choice of winners.
In the sequel, we will construct characteristic functions mostly using tournament matrices.
It is desired that the existing procedures of ballot paper handling and winner definition
meet a series of properties.

{\bf Unanimity.} If each voter ranks candidate $x$ not lower than candidate $y$,
then the former appears not lower than the latter in the collective preference.

Candidate $x$ is termed the Condorcet winner if $x$ beats any other candidate
in the case of pairwise comparison. This means that for any $y\in A\setminus \{x\}$
over one-half of the voters rank $x$ higher than $y$, i.e., $h(x,y)>n/2$.

{\bf The Condorcet property.} If a candidate is the Condorcet winner, then
it ranks first in the collective preference.

{\bf Monotonicity.} Suppose that in its individual preference a voter moves candidate $x$
by one position up (down) under a fixed ranking of all other candidates;
then in the collective preference $x$ does not decrease (increase, respectively) its rank.

These properties will be supplemented by a couple of new ones
for the ranking procedures suggested below.

Denote by $w(x)$ and $l(x)$ the sets of candidates loosing to and winning against candidate $x$,
respectively, in the case of pairwise comparison. Interestingly, candidate $x$
represents the Condorcet winner iff $w(x)=A\setminus \{x\}$ and $l(x)=\emptyset$.

{\bf The strong Condorcet property.} If $w(x)\supseteq w(y)$, $l(x)\subseteq l(y)$
and $h(x,y)>n/2$, then in the collective preference candidate $x$ ranks not lower
than candidate $y$.

{\bf The majority rule.} For any initial preference profile and any order of candidates
$a_1>a_2>...>a_m$, there exists an integer number $N$ such that after adding $k\geq N$
ballot papers with the given order of candidates the new preference profile
yields a collective ranking coinciding with the ranking in the group of added ballot papers.

\noindent {\bf Lemma~1.} \textit{ A ranking procedure meeting the strong Condorcet property
also satisfies the majority rule.}

To prove Lemma~1, it suffices to observe the following fact. Adding identical ballot papers
with the order $a_1>a_2>...>a_m$, whose number exceeds by 1 the initial number of ballot papers,
brings to the collective ranking $a_1>a_2>...>a_m$.

In the forthcoming sections, we will check these properties for a new ranking procedure.
This procedure involves a certain cooperative game associated with
the filled ballot paper profile.

Let $K\subseteq A$ indicate a coalition of candidates. Suppose that each coalition $K$
is assigned with a nonnegative monotonous function $v(K)$ (a characteristic function
in the terminology of cooperative game theory). Find the characteristic function $v(K)$
for a given voter preference profile. Then candidate ranking can be performed
on the basis of cooperative game theory criteria adopted in voting problems analysis.
In the sequel, the role of such criterion belongs to the candidate's power in the form
of the Shapley value. In this case, candidate ranking runs according to
the Shapley values for a given characteristic function. First, we will define
characteristic functions using tournament matrices.

\bigskip

{\large \bf 3. Characteristic function as the value of a constant-sum game}

Define the characteristic function as follows. Consider a coalition $K$
and its complement $A\setminus K$. Assume that the coalition $K$
proposes for elections a common candidate $i\in K$, whereas
the coalition $A\setminus K$ nominates its representative $j\in A\setminus K$.
The candidate receiving over half of the votes becomes the winner;
otherwise, the elections are drawn. The payoff in this game makes up
\begin{equation*}
H(i,j)=I\left(h(i,j)-\frac{n}{2}\right),
\end{equation*}
where the indicator function $I(z)=1$ for $z>0$, $I(z)=1/2$ for $z=0$ and $0$
for the rest values of $z$.

This mixed strategy game has an equilibrium according to the Nash theorem;
moreover, its value gives the value of the characteristic function $v(K)$
in the cooperative game.

Therefore, the payoff $v(K)$ of the coalition $K$ makes the equilibrium payoff
in the constant-sum game of the coalition $K$ against the countercoalition $A\setminus K$.
A mixed strategy of the coalition $K$ is a vector $p=(p_i)_{i\in K}$. The coalition $K$
proposes its common candidate $i\in K$ with the probability $p_i\geq 0$,
where $\sum\limits_{i\in K}{p_i}=1.$ A strategy of the coalition $A\setminus K$
is a vector $q=(q_j)_{j\in A\setminus K}$ such that $q_j\geq 0$ for all $j\in
A\setminus K$, where $\sum\limits_{j\in A\setminus K}{q_j}=1.$
Then the characteristic function $v$ acquires the form
\begin{equation*}
v(K)=\max_{p} {\min_{q} {\sum\limits_{i\in K}{ \sum\limits_{j\in
A\setminus K}{ H(i,j)p_{i}q_{j}}}}}.
\end{equation*}
Note that
\begin{equation*}
v(K) + v(A\setminus K)=1.
\end{equation*}

For instance, calculate the payoff $v$ for the coalitions $ac$ and $bde$
in Example~1. The payoff matrix of the coalition $ac$ against the coalition $bde$
is defined by

$$
 \bordermatrix{
    & b & d & e \cr
  a & 0 & 1 & 0 \cr
  c & 1 & 0 & 1 \cr}.
$$

In mixed strategies, the coalition $ac$ has the payoff $0.5$;
hence, $v(ac)=v(bde)=0.5$.

A coalition wins at least by 23 affirmative votes. The coalitions $ce, abe$
and $acd$ are minimal winning coalitions. Candidates $c$ and $e$
possess the highest power under such voting procedure. Any winning coalition
must include candidate $c$ or $e$.

The characteristic function can be calculated for all $2^m$ coalitions of
candidates. After construction of the characteristic function,
it is possible to evaluate the candidate's power using the Shapley value:
\begin{equation*}
\varphi_x(v)=\sum_{K:x\not\in K}\frac{k!(m-k-1)!}{m!} \left(v(K\cup
x)-v(K)\right), \quad x\in A.
\end{equation*}

The Appendix provides the corresponding table with all values for
the characteristic function $v$ in Example~1.

\begin{center}
{\small\textit{The Shapley value for the characteristic function in Example~1.} \\
\vspace{3mm}
\begin{tabular}{|c|c|c|c|c|c|c|}
\hline
  & $a$ & $b$ & $c$ & $d$ & $e$ & ranking \\
\hline
$v$ &  0.167 & 0.083 & 0.333 & 0.083 & 0.333 &  $e=c>a>b=d$\\
\hline
\end{tabular}
}
\end{center}

Further exposition employs two auxiliary lemmas.

\noindent {\bf Lemma~2.} \textit{ Let $(h_{i j})$ be a payoff matrix
of dimensions $n\times m$ in the constant-sum game with a value $v^*$.
Then the game with a matrix $\widehat{h}$ such that
$\widehat{h}_{i j}\geq h_{i j}$ for all $i, j$ has a value not smaller than $v^*$.}

\textsl{ Proof.}

Designate by $p^*$ and $q^*$ any optimal strategies of players 1 and 2
in the game with the matrix $h$. It suffices to show that in the game with
the matrix $\widehat{h}$ the strategy $p^*$ guarantees to player 1
a payoff not less than $v^*$. Imagine that there exists a strategy $q$
of player 2 such that the payoff of its opponent is smaller than $v^*$.
In this case,
\begin{equation*}
\sum\limits_{i=1}^n{ \sum\limits_{j=1}^m { p_i^* q_j \widehat{h}_{i
j} } }<v^*\leq \sum\limits_{i=1}^n{ \sum\limits_{j=1}^m { p_i^* q_j
h_{i j} } },
\end{equation*}
which is impossible under $\widehat{h}_{i j}\geq h_{i j}$.

\noindent {\bf Lemma~3.} \textit{Suppose that for some pair of candidates $x, y$
and any coalition $S\subseteq A\setminus \{x, y\}$ the characteristic functions
$v$ and $\widehat{v}$ meet the following conditions:
\begin{equation*}
v(S)=\widehat{v}(S),
\end{equation*}
\begin{equation*}
v(S\cup y)\geq \widehat{v}(S\cup y),
\end{equation*}
\begin{equation*}
v(S\cup x)\leq \widehat{v}(S\cup x),
\end{equation*}
\begin{equation*}
v(S\cup y \cup x) = \widehat{v}(S\cup y \cup x).
\end{equation*}}

\textit{ Then transition from $v$ to $\widehat{v}$ does not
decrease the Shapley value $\varphi_x$ for candidate $x$
and does not increase the Shapley value $\varphi_y$ for candidate $y$.
For any other candidate $z\in A\setminus \{x, y\}$,
the Shapley value increment $\varphi_z(\widehat{v})-\varphi_z(v)$
is not greater than the increment $\varphi_x(\widehat{v})-\varphi_x(v)$ for $x$
and not smaller than the increment $\varphi_y(\widehat{v})-\varphi_y(v)$ for $y$.
Consequently, candidate $x$ does not decrease its rank and candidate $y$
does not increase its rank in the collective preference as the result of
replacing the characteristic function $v$ by $\widehat{v}$.}

\textsl{ Proof.}

The premises of the lemma directly imply that the Shapley value does not
decrease (increase) for candidate $x$ (for candidate $y$, respectively).
Now, demonstrate that for any other candidate $z\in A\setminus \{x, y\}$
the Shapley value increment does not exceed that for $x$.

For this, estimate the quantities
\begin{equation}
\varphi_x(\widehat{v}){-}\varphi_x(v){=} \sum_{K:x\not\in
K}\frac{k!(m{-}k{-}1)!}{m!} (\widehat{v}(K\cup
x){-}\widehat{v}(K)){-}(v(K\cup x){-}v(K))
\end{equation}
and
\begin{equation}
\varphi_z(\widehat{v}){-}\varphi_z(v){=} \sum_{K:z\not\in
K}\frac{k!(m{-}k{-}1)!}{m!} (\widehat{v}(K\cup
z){-}\widehat{v}(K)){-}(v(K\cup z){-}v(K))).
\end{equation}
Introduce the notation
$$
\Delta_K(x)=(\widehat{v}(K\cup x)-\widehat{v}(K))-(v(K\cup x)-v(K)).
$$

If $x,z \not\in K$, then for $y\not\in K$ we have

\begin{equation*}
\widehat{v}(x\cup K) - \widehat{v}(K)-(v(x\cup
K)-v(K))=\widehat{v}(x\cup K)-v(x\cup K)\geq 0.
\end{equation*}
At the same time,
\begin{equation*}
\widehat{v}(z\cup K) - \widehat{v}(K)-(v(z\cup K)-v(K))= 0.
\end{equation*}
This dictates that $\Delta_K(x)\ge \Delta_K(z)$. On the other hand, if $y \in K$,
the inequalities
\begin{equation*}
\widehat{v}(x\cup K) - \widehat{v}(K)-(v(x\cup K)-v(K))=
v(K)-\widehat{v}(K)\geq 0,
\end{equation*}
$$
\widehat{v}(z\cup K) - \widehat{v}(K)-(v(z\cup K)-v(K))=
$$
$$
=\widehat{v}(z\cup y\cup (K\setminus y))-v(z\cup y\cup (K\setminus y))- (\widehat{v}(K)-v(K))
\leq v(K)-\widehat{v}(K),
$$
again yield $\Delta_K(x)\ge \Delta_K(z)$.

Next, consider the coalitions $K$ in the sum~(1), which contain player $z$.
Associate them with the coalitions $K'=x \cup (K\setminus z)$ in the sum~(2).
The inequalities below hold true depending on whether player $y$
belongs to this coalition or not.

If $y\not\in K$, then
$$ \Delta_x(K)=
\widehat{v}(x\cup K) - \widehat{v}(K)-(v(x\cup K)-v(K))=
$$
$$=\widehat{v}(x\cup K)-v(x\cup K))\geq 0,
$$
whereas
$$
\Delta_z(K')=
\widehat{v}(z\cup  K') - \widehat{v}(K')-(v(z\cup K')-v(K'))=
$$
$$
=\widehat{v}(z\cup x \cup (K\setminus z)) - \widehat{v}(x \cup (K\setminus z))-(v(z\cup x \cup (K\setminus z))-v(x \cup (K\setminus z)))\leq
$$
$$\leq \widehat{v}(x\cup K)-v(x\cup K)=\Delta_x(K).
$$
In the case of $y \in K$, we obtain
$$
\Delta_x(K)=
\widehat{v}(x\cup y\cup  (K\setminus y)) {-} \widehat{v}(y\cup (K\setminus y)){-}(v(x\cup y\cup (K\setminus y)){-}v(y\cup (K\setminus y))=
$$
$$=
v(y\cup (K\setminus y)){-}\widehat{v}(y\cup (K\setminus y)){\geq} 0,
$$
whereas
$$
\Delta_z(K')=
\widehat{v}(z\cup  K') - \widehat{v}(K')-(v(z\cup K')-v(K'))=
$$
$$
=
\widehat{v}(x \cup K) - \widehat{v}(x \cup (K\setminus z))-(v(x \cup K))-v(x \cup (K\setminus z)))=0.
$$
Hence, it appears that $\Delta_x(K)\geq \Delta_z(K')$. Thus, we have established that
$$
\varphi_x(\widehat{v})-\varphi_x(v)\geq \varphi_z(\widehat{v})-\varphi_z(v).
$$

Similarly, it is possible to demonstrate that that the Shapley value increment for $y$
does not exceed that for $z$. Summarizing the outcomes, we have argued that
candidate $x$ (candidate $y$) does not decrease (increase, respectively) its rank
in the collective preference as the characteristic function $v$ is replaced by $\widehat{v}$.

\noindent {\bf Theorem~1.} \textit{ The characteristic function $v$
is nonnegative and monotonous. Ranking based on the Shapley value
for the function $v$ enjoys the properties of unanimity and monotonicity,
the majority rule, the Condorcet and strong Condorcet properties.}

\textsl{ Proof.}
The nonnegativity and monotonicity of the function $v$ is obvious from
the definition. Note that, under an odd number of voters, this function
possesses the superadditive property.

First, we show the property of unanimity. Suppose that candidate $x$ is
preferable to candidate $y$ for all voters. It suffices to verify
the inequality $v(y\cup S)\leq v(x\cup S)$ for any coalition
$S\subseteq A\setminus \{x, y\}$. Denote $K=A\setminus \{x, y\}\setminus S$.
Interestingly, $h(i,x)\leq h(i,y)$ and $h(y,j)\leq h(x,j)$ for any $i, j$.
This means that $H(i,x)\leq H(i,y)$ and $H(y,j)\leq H(x,j)$ for any $i, j$.
Compare the payoff matrix of the coalition $y\cup S$ against $x\cup K$

\begin{equation}
\bordermatrix{
 & x & k_1 & \ldots & k_r \cr
y & 0 & H(y,k_1) & \ldots & H(y,k_r) \cr
 s_1 & H(s_1,x) & H(s_1,k_1) & \ldots & H(s_1, k_r) \cr
\ldots & \ldots & \ldots & \ldots & \ldots \cr s_l & H(s_l,x) &
H(s_l,k_1) & \ldots & H(s_l, k_r)}
\end{equation}

with the payoff matrix of the coalition $x\cup S$ against $y\cup K$

\begin{equation}
\bordermatrix{
 & y & k_1 & \ldots & k_r \cr
x & 1 & H(x,k_1) & \ldots & H(x,k_r) \cr s_1 & H(s_1,y) & H(s_1,k_1)
& \ldots & H(s_1, k_r) \cr \ldots & \ldots & \ldots & \ldots &
\ldots \cr s_l & H(s_l,y) & H(s_l,k_1) & \ldots & H(s_l, k_r)}
\end{equation}

Clearly, elements in the lower matrix are not smaller than their counterparts
in the upper matrix. By virtue of Lemma~2, we have $v(y\cup S)\leq v(x\cup S)$.
And so, the Shapley value for candidate $x$ is not less than that for candidate $y$.

Next, let us prove the monotonicity of the ranking procedure. Assume that in a ballot paper
candidate $x$ moves by one position up, whereas candidate $y$ goes by one position down.
Designate by $\widehat{v}$ the characteristic function resulting from such transformation.
Obviously, for any coalition $S\subseteq A\setminus \{x, y\}$
the conditions of Lemma~3 take place. According to Lemma~3, candidate $x$ (candidate $y$)
does not decrease (increase, respectively) its rank in the collective preference.

Third, check the Condorcet property. Imagine that candidate $x$ represents
the Condorcet winner and compare it with any other candidate $y$.
For any set $S\subseteq A\setminus \{x, y\}$, the coalition $x\cup S$
proposes the common candidate $x$ and wins. Consequently, we have
\begin{equation*}
1=v(x\cup S)> v(y\cup S)=0,
\end{equation*}
whence it follows that the Shapley value is higher for candidate $x$ than for candidate $y$.

And finally, establish the strong Condorcet property. Let $w(x)\supseteq w(y)$,
$l(x)\subseteq l(y)$ and $h(x,y)>n/2$. Then $I(h(i,x)-\frac{n}{2})\leq I(h(i,y)-\frac{n}{2})$,
$I(h(y,j)-\frac{n}{2})\leq I(h(x,j)-\frac{n}{2})$ for any $i, j$.
Verify the inequality $v(y\cup S)\leq v(x\cup S)$ for any set
$S\subseteq A\setminus \{x, y\}$. Designate $K=A\setminus \{x, y\}\setminus S$.
The payoff matrices of the coalition $y\cup S$ against $x\cup K$
coincide with the matrices~(3) and (4), see the proof of unanimity.
Then Lemma~2 brings to the condition $v(y\cup S)\leq v(x\cup S)$.
Consequently, for candidate $x$ the Shapley value is not less than
for candidate $y$.

The function $v$ satisfies the strong Condorcet property, \textit{ergo}
the majority rule (see Lemma~1). This concludes the proof of Theorem~1.

The characteristic function $v$ takes into account merely the win
of one candidate over another under pairwise comparison. Here
the advantage of one vote and unanimity are equivalent.
Such ranking method reflects candidate's capability for
creating coalitions that propose the Condorcet winner.
If the Condorcet winner is among all candidates,
the Shapley value vector makes up $(1,0,\ldots,0)$.

{\bf Remark}. We have defined the value of the characteristic function
as the value of a constant-sum game with mixed strategies. An alternative approach
is to introduce the characteristic function in terms of pure strategies only.
Then its value in the game of a coalition $K$ against the countercoalition
$A\setminus K$ becomes
\begin{equation*}
v(K)=\max_{i\in K} {\min_{j\in A\setminus K} {H(i,j)}}.
\end{equation*}
Note that Theorem~1 remains in force for such characteristic function, either.

\bigskip

{\large \bf 4. Ranking based on tournament matrix}

To calculate the function $v$, we have utilized the payoff matrix
composed of zeros and unities. In what follows, let us estimate
the advantage of one candidate over another in a more accurate way.
As previously, consider the constant-sum game of a coalition $K$
against the countercoalition $A\setminus K$. Each coalition proposes
a common candidate. By assumption, the payoff of a coalition
is the number of votes $h(i,j)$ polled by common candidate $i\in K$
against common candidate $j\in A\setminus K$.

For the coalition $K$, the payoff $u(K)$ is its equilibrium payoff
in the game against the countercoalition $A\setminus K$.
In this case, the optimal strategy of the coalition $K$ (i.e., choosing
common candidate $i \in K$) may appear mixed. A mixed strategy
of the coalition $K$ represents a vector $p=(p_i)_{i\in K}$. A strategy
of the coalition $A\setminus K$ forms a vector $q=(q_j)_{j\in A\setminus K}$.
Therefore, the characteristic function is defined by
\begin{equation*}
u(K)=\max_{p} {\min_{q} {\sum\limits_{i\in K}{ \sum\limits_{j\in
A\setminus K}{ h(i,j)p_{i}q_{j}}}}}.
\end{equation*}
For any coalition $K$, we have the equality
\begin{equation*}
u(K) + u(A\setminus K)=n.
\end{equation*}

For instance, revert to Example~1 and find the values of the function $u$
for the coalitions $ac$ and $bde$. The payoff matrix of the coalition $ac$
against the coalition $bde$ takes the form

$$
 \bordermatrix{
 & b & d & e \cr
a & 20 & 30 & 22  \cr
 c & 29 & 17 & 24}.
$$

For the coalition $ac$, common candidate $a$ guarantees 20 votes against $b$,
whereas candidate $c$ ensures 17 votes against $d$. For the coalition $bde$,
common candidates $b$, $d$ and $e$ guarantee 45-29=16 votes against $c$,
45-30=15 votes against $a$ and 45-24=21 votes against $c$, respectively.
Here the guaranteed payoffs under mixed strategies constitute $u(ac)=20$ and
$u(bde)=45-24=21$. In the mixed strategy equilibrium, the coalition $ac$
uses probabilities for the strategies $(7/15,8/15)$, and the coalition $bde$
does same for the strategies $(0,2/15,13/15)$. The corresponding payoffs
are $u(ac)=346/15\approx 23.07$ and $u(bde)=329/15\approx 21.93$.

For the sake of comparison, we explicitly define other characteristic functions
$v_{1}, v_2, v_{3}, v_4$ by the formulas
\begin{equation*}
v_1(K)=\max_{i\in K} \min_{j\in A\setminus K} h(i,j),
\end{equation*}
\begin{equation*}
v_{2}(K)=\sum_{i\in K} {\min_{j\in A\setminus K} {H(i,j)}},
\end{equation*}
\begin{equation*}
v_{3}(K)=\sum_{i\in K} {\min_{j\in A\setminus K} {h(i,j) H(i,j)}},
\end{equation*}
\begin{equation*}
v_{4}(K)=\sum_{i\in K} {\min_{j\in A\setminus K} {h(i,j)}}.
\end{equation*}

The table with all values of the characteristic functions $u, v_{1}-v_{4}$
in Example~1 can be found in the Appendix.

Theorem~2 below studies the properties of the Shapley value-based ranking procedure
for the characteristic functions $u, v_{1}-v_{4}$. For a rigorous proof, we refer
an interested reader to the Appendix.

\noindent {\bf Theorem~2.} \textit{ The characteristic functions $u$, $v_{1}-v_4$
are nonnegative and monotonous, and the functions $v_{2} - v_{4}$ also enjoy
superadditivity. The Shapley value-based ranking procedure for the functions
$u, v_1-v_{4}$ possesses the properties presented in the table below.}
\begin{center}
{\small\textit{Satisfaction of the Shapley value properties
by the characteristic functions $u, v_1-v_{4}$.} \\
\vspace{3mm}
\begin{tabular}{|c|c|c|c|c|c|c|c|c|c|c|c|}
\hline
Property  & $u$ & $v_1$ & $v_2$ &  $v_{3}$ & $v_{4}$ & Borda & Copeland & Maximin \\
\hline
Unanimity & yes  & yes  & yes  &  yes & yes & yes & yes & yes \\
\hline
Monotonicity & yes  & yes  & yes  &  yes & yes  & yes & yes & yes \\
\hline
Majority rule & no  & no  & yes  & yes &  yes & yes & yes & no \\
\hline
Condorcet & yes  & yes  & yes  &  yes & no & no & yes & yes \\
\hline
Strong Condorcet & no  & no  & yes  & no & no & no & yes & no \\
\hline
\end{tabular}
}
\end{center}

Compare the results derived in Example~1 with other ranking and
winner definition methods involving tournament matrices (see Klamler
2006). A well-known ranking technique of $m$ candidates is the Borda
rule: a candidate receives $m-1$ points for rank 1, $m-2$ points for
rank 2, \ldots, 0 points for rank $m$ in a ballot paper. The winner
is the candidate having the maximum total points. For candidate $i$,
the total points have the form
\begin{equation*}
\sum_{j\in A\setminus \{ i \}} h(i,j).
\end{equation*}
In Example~1, the Borda rule yields the ranking $e>a>b>c>d$
with the total points of 102, 98, 92, 89 and 69, respectively.

The Copeland method proceeds from points calculated by
\begin{equation*}
\sum_{j\in A\setminus \{ i \}} H(i,j).
\end{equation*}
In our case, the method brings to the ranking $e>a=b=c>d$ with the total points of
3, 2, 2, 2 and 1, respectively.

According to the maximin rule, the total points of candidate $i$ are defined by
\begin{equation*}
\min_{j\in A\setminus \{ i \}} h(i,j).
\end{equation*}
Clearly, we obtain the ranking $e>a>c>b>d$ with the total points of 21, 20, 17, 16
and 12, respectively.

And the Schulze method leads to the ranking $e>a>c>b>d$.

\begin{center}
{\small\textit{ The Shapley values of the characteristic functions in Example~1.} \\
\vspace{3mm}
\begin{tabular}{|c|c|c|c|c|c|c|}
\hline
  & $a$ & $b$ & $c$ & $d$ & $e$ & ranking \\
\hline
$u$ &  10.994 &  7.9 & 9.161 & 5.306 & 11.639 &  $e>a>c>b>d$\\
\hline
$v_1$ & 10.95 & 7.867 & 9.033 & 5.367 & 11.783 & $e>a>c>b>d$\\
\hline
$v_{2}$ & 0.917 & 0.917 & 1.167 & 0.583 & 1.417 &  $e>c>a=b>d$\\
\hline
$v_{3}$ & 45.05 & 42.8 & 49.217 & 31.05 & 56.883 &  $e>c>a>b>d$\\
\hline
$v_{4}$ &  48.633 & 45.05 & 45.383 & 35.3 & 50.633 &  $e>a>c>b>d$\\
\hline
Borda & 98 & 92 & 89 & 69 & 102 &  $e>a>b>c>d$\\
\hline
Copeland & 2 & 2 & 2 & 1 & 3 &  $e>a=b=c>d$\\
\hline
Maximin & 20 & 16 & 17 & 12 & 21 &  $e>a>c>b>d$\\
\hline
\end{tabular}
}
\end{center}

{\bf Example~2.} Consider a situation with $m=3$ candidates and $n=2k+1$ voters.
Imagine that $k+1$ voters choose the ranking $a>b>c$,
whereas the rest $k\geq 1$ ones prefer $b>c>a$.
\begin{center}
{\small\textit{The preference profile in Example~2.} \\
\vspace{3mm}
\begin{tabular}{|c|c|c|c|c|c|c|c|}
\hline
$k+1$ & $k$  \\
\hline
a & b  \\
b & c  \\
c & a  \\
\hline
\end{tabular}
}
\end{center}
Candidate $a$ is the Condorcet winner; however, it beats $b$ and $c$ merely
by the advantage of one vote. Candidate $b$ seems preferable to $c$ for all voters.
The pairwise comparison of all candidates leads to the ranking $a>b>c$ as the collective decision.
\begin{center}
{\small\textit{The tournament matrix in Example~2.} \\
\vspace{3mm}
\begin{tabular}{|c|c|c|c|c|c|c|c|}
\hline
  & a & b & c  \\
\hline
a &   & $k+1$ & $k+1$  \\
\hline
b & $k$ &   & $2k+1$   \\
\hline
c & $k$ & 0 &    \\
\hline
\end{tabular}
}
\end{center}
The table combines the Shapley values for the functions $u, v_{1}-v_{4}$.
This example demonstrates that $v_{4}$ does not obey the Condorcet condition.
\begin{center}
{\small\textit{ The Shapley values of the characteristic functions in Example~2.} \\
\vspace{3mm}
\begin{tabular}{|c|c|c|c|c|c|c|c|c|c|}
\hline
  &  $u$=$v_{1}$ & $v_{2}$ & $v_{3}$ & $v_{4}$ & Borda \\
\hline
$a$ &  $k+1$  & $11/6$ & $3k+11/6$ &  $\frac{13}{6}k+\frac{11}{6}$ & $2k+2$ \\
\hline
$b$ &  $k$  & $5/6$ & $2k+5/6$ &  $\frac{16}{6}k+\frac{5}{6}$ & $3k+1$ \\
\hline
$c$ &  0  & 1/3 & $k+1/3$ &  $\frac{7}{6}k+\frac{1}{3}$ & $k$ \\
\hline
\end{tabular}
}
\end{center}

{\bf Example~3.} Consider the case of $m=3$ candidates and $n=k+1$ voters.
Suppose that $k\geq 2$ voters choose the ranking $a>b>c$, and one voter prefers $c>a>b$.
 \begin{center}
{\small\textit{The preference profile in Example~3.} \\
\vspace{3mm}
\begin{tabular}{|c|c|}
\hline
$k$ & $1$  \\
\hline
a & c  \\
b & a  \\
c & b  \\
\hline
\end{tabular}
}
\end{center}
The pairwise comparison of all candidates leads to the unique collective ranking $a>b>c$.
\begin{center}
{\small\textit{The tournament matrix in Example~3.} \\
\vspace{3mm}
\begin{tabular}{|c|c|c|c|}
\hline
  & a & b & c  \\
\hline
a &   & $k+1$ & $k$  \\
\hline
b & $0$ &   & $k$   \\
\hline
c & $1$ & 1 &    \\
\hline
\end{tabular}
}
\end{center}
The table presents the Shapley values for the characteristic functions $u, v_{1}-v_{4}$.
This example shows that the functions $u$ and $v_{1}$ do not meet the majority rule.
\begin{center}
{\small\textit{ The Shapley values of the characteristic functions in Example~3.} \\
\vspace{3mm}
\begin{tabular}{|c|c|c|c|c|c|c|c|c|}
\hline
  &   $u$=$v_{1}$ & $v_{2}$ & $v_{3}$ & $v_{4}$ & Borda \\
\hline
$a$ &  $k$  & $11/6$ & $\frac{11}{6}k+\frac{7}{6}$ &  $\frac{11}{6}k+\frac{5}{6}$ & $2k+1$ \\
\hline
$b$ &  0  & $5/6$ & $\frac{5}{6}k+\frac{2}{3}$ &  $\frac{5}{6}k+\frac{1}{3}$ & $k$ \\
\hline
$c$ &   1  & 1/3 & $\frac{1}{3}k+\frac{7}{6}$ &  $\frac{1}{3}k+\frac{11}{6}$ & 2 \\
\hline
\end{tabular}
}
\end{center}

{\bf Example~4.} There are $m=5$ candidates and $n=7$ voters.
Four of them choose the ranking $a>b>c>d>e$, and the rest three voters prefer $e>a>b>c>d$.
 \begin{center}
{\small\textit{The preference profile in Example~4.} \\
\vspace{3mm}
\begin{tabular}{|c|c|c|c|c|c|c|c|}
\hline
$4$ & $3$  \\
\hline
a & e  \\
b & a  \\
c & b  \\
d & c  \\
e & d  \\
 \hline
\end{tabular}
}
\end{center}
Adhering to the strong Condorcet condition brings to the collective ranking $a>b>c>d>e$
and candidate $e$ receives rank 5 (really, it loses to all opponents under pairwise
comparison). Nevertheless, almost half of the voters assign rank 1 to it.
Below, we will demonstrate that candidate $e$ has higher ranks in other ranking procedures.
\begin{center}
{\small\textit{The tournament matrix in Example~4.} \\
\vspace{3mm}
\begin{tabular}{|c|c|c|c|c|c|c|c|}
\hline
  & a & b & c & d & e  \\
\hline
a &   & 7 & 7 & 7 & 4  \\
\hline
b & 0 &   & 7 & 7 & 4  \\
\hline
c & 0 & 0 &  & 7 & 4  \\
\hline
d & 0 & 0 & 0 &   & 4  \\
\hline
e & 3 & 3 & 3 & 3 &    \\
\hline
\end{tabular}
}
\end{center}
Example~4 shows that the functions $u$, $v_{1}$, $v_{3}$, and $v_{4}$
do not satisfy the strong Condorcet condition.
\begin{center}
{\small\textit{ The Shapley values of the characteristic functions in Example~4.} \\
\vspace{3mm}
\begin{tabular}{|c|c|c|c|c|c|c|}
\hline
  & $a$ & $b$ & $c$ & $d$ & $e$ & ranking \\
\hline
$u=v_{1}$ & 4 & 0 & 0 & 0 & 3 & $a>e>b=c=d$\\
\hline
$v_{2}$ & 2.283 & 1.283 & 0.783 & 0.45 & 0.2 & $a>b>c>d>e$\\
\hline
$v_{3}$ & 13.583 & 8.083 & 5.083 & 3 & 5.25 &  $a>b>e>c>d$\\
\hline
$v_{4}$ & 12.983 & 7.483 & 4.483 & 2.4 & 7.65 & $a>e>b>c>d$\\
\hline
Borda & 25 & 18 & 11 & 4 & 12 & $a>b>e>c>d$\\
\hline
\end{tabular}
}
\end{center}

{\bf Example~5.} To illustrate the obtained results, we consider
2010 FIA Formula One World Championship. Actually, 27 drivers participated
in the championship and 19 of them gained points. Five drivers competed for the champion's title.
The championship consisted of 19 races (Grand-Prix), where 10 best drivers earned
25, 18, 15, 12, 10, 8, 6, 4, 2, and 1 point(s) for the positions from 1 to 10, respectively.
Denote by \textit{PS1} this point system. In the individual event, the points
gained at all 19 races were summed up. During 2003--2009 seasons, points were assigned
to the first 8 drivers in the final classification of a race according to
the following point system (\textit{PS2}): 10, 8, 6, 5, 4, 3, 2, and 1 point(s).
The period from 1991 to 2002 was remarkable for another point system (\textit{PS3}):
10, 6, 4, 3, 2, and 1 point(s). For the sake of comparison, the table presents
the points in the individual event of the championship counted by these three systems.
\begin{center}
{\small\textit{ 2010 FIA Formula One World Championship in the individual event.} \\
\vspace{3mm}
\begin{tabular}{|c|c|c|c|c|c|c|}
\hline
&Fernando&Sebastian&Lewis(L)&Jenson&Mark(M)&Individual\\
&Alonso(F)&Vettel(S)&Hamilton&Button(D)&Webber&event\\
\hline
PS1& 252 & 256 & 240 & 214 & 242 &S{>}F{>}M{>}L{>}D\\
\hline
PS2& 101 & 104 & 100 & 87 & 97 &S{>}F{>}L{>}M{>}D\\
\hline
PS3& 81 & 84 & 76 & 61 & 76 &S{>}F{>}M{>}L{>}D\\
\hline
\end{tabular}
}
\end{center}

Clearly, L.~Hamilton would occupy the third position in 2010 FIA Formula One
World Championship instead of M.~Webber if the organizers preserved
the same point system as in 2009.

The ranking problem of 27 drivers requires bulky computations.
Therefore, let us consider the relative results of the five leading drivers
in the championship.
\begin{center}
{\small\textit{The relative classification of the five drivers at different races.} \\
\vspace{3mm}
\begin{tabular}{|c|c|c|c|c|c|c|c|c|c|c|c|c|c|c|c|c|c|c|}
\hline
 1 & 2 & 3 & 4 & 5 & 6 & 7 & 8 & 9 & 10 & 11 & 12 & 13 & 14 & 15 & 16 & 17 & 18 & 19 \\
\hline
 F & D & S & D & M & M & L & L & S & M & F & M & L & F & F & S & F & S & S \\
 L & F & M & L & F & S & D & D & L & L & S & F & M & D & S & M & L & M & L \\
 S & L & L & F & S & L & M & F & D & D & L & S & S & S & M & F & D & F & D \\
 D & M & D & S & D & F & F & S & F & S & D & D & F & M & D & D & S & L & F \\
 M & S & F & M & L & D & S & M & M & F & M & L & D & L & L & L & M & D & M \\
\hline
\end{tabular}
}
\end{center}

\begin{center}
{\small\textit{The tournament matrix for the five drivers.} \\
\vspace{3mm}
\begin{tabular}{|c|c|c|c|c|c|c|c|}
\hline
  & F & S & L & D & M  \\
\hline
F &   & 11 & 10 & 11 & 10  \\
\hline
S & 8 &   & 11 & 12 & 12  \\
\hline
L & 9 & 8 &  & 12 & 10  \\
\hline
D & 8 & 7 & 7 &   & 10  \\
\hline
M & 9 & 7 & 9 & 9 &    \\
\hline
\end{tabular}
}
\end{center}

Ranking of the drivers based on pairwise comparison yields a unique result: F>S>L>D>M.

\begin{center}
{\small\textit{The relative rank distribution of the five drivers.} \\
\vspace{3mm}
\begin{tabular}{|c|c|c|c|c|c|c|c|}
\hline
Rank  & F & S & L & D & M  \\
\hline
1 &  5 & 5 & 3 & 2 & 4  \\
2 & 3 & 3  & 6 & 3 & 4  \\
3 & 4 & 5 & 4 & 4 & 2  \\
4 & 5 & 4 & 1 & 7  & 2  \\
5 & 2 & 2 & 5 & 3 &  7  \\
\hline
\end{tabular}
}
\end{center}

Interestingly, the rankings S>F>D and S>F>M result from any point system,
see the table below.

\begin{center}
{\small\textit{ The results shown by the five drivers: a comparison.} \\
\vspace{3mm}
\begin{tabular}{|c|c|c|c|c|c|c|}
\hline
& F & S & L & D & M & Ranking\\
\hline
PS1 & 319 & 322 & 305 & 278 & 296 &S{>}F{>}L{>}M{>}D\\
\hline
PS2 & 131 & 132 & 127 & 115 & 122 &S{>}F{>}L{>}M{>}D\\
\hline
PS3 & 103 & 104 & 95 & 81 & 92 &S{>}F{>}L{>}M{>}D\\
\hline
Borda & 42 & 43 & 39 & 32 & 34 &S{>}F{>}L{>}M{>}D\\
\hline
$v$ & 1 & 0 & 0 & 0 & 0 &F{>}S{=}L{=}D{=}M\\
\hline
$u$=$v_{1}$ & 5.633 & 3.967 & 3.633 & 2.8 & 2.967 &F{>}S{>}L{>}M{>}D\\
\hline
$v_{2}$ & 2.283 & 1.283 & 0.783 & 0.45 & 0.2 &F{>}S{>}L{>}D{>}M\\
\hline
$v_{3}$ & 32.5 & 22.167 & 17 & 12.5 & 10.833 &F{>}S{>}L{>}D{>}M\\
\hline
$v_{4}$ & 21.7 & 21.2 & 19.033 & 16.033 & 17.033 &F{>}S{>}L{>}M{>}D\\
\hline
\end{tabular}
}
\end{center}

{\large \bf 5. Conclusion}.

This paper has employed cooperative game theory methods to solve
the ranking problem of candidates for a certain position.
For this, it is necessary to construct a characteristic function
using the filled ballot papers of voters; such function defines
the payoff of each coalition. The next step is to find the Shapley value
which serves as the ranking method. Note that the stated ranking procedure
takes into account the weight of each candidate in all possible coalitions.
This feature allows applying the ranking procedure for defining
a single winner or several winners (the composition of a certain committee).
And finally, the paper has compared this method with other well-known
candidate ranking procedures.

This research is supported by the Russian Fund for Basic Research
(project 13-01-00033-a) and the Division of Mathematical Sciences of
Russian Academy of Sciences.

{\bf References}.

1. Brams SJ, Fishburn PC (1978) Approval Voting. \textit{The
American Political Science Review}, vol.~72, no.~3. Pp.~831--847.

2. Brams SJ, Fishburn PC (2005) Going from Theory to Practice: The
Mixed Success of Approval Voting. \textit{Social Choice and
Welfare}, vol.~25, no.~2--3. Pp.~457--474.

3. Hillinger C (2005) The Case for Utilitarian Voting, \textit{Homo
Oeconomicus}, vol.~23. Pp.~295--321.

4. Gaertner W, Xu Y (2012) A General Scoring Rule,
\textit{Mathematical Social Sciences}, vol.~63, no.~3. Pp.~193--196.

5. Smith WD (2000) Range Voting, \textit{Technical Report 56}, NEC
Research, Princeton, NJ, USA.

6. Balinski M,  Laraki R (2007) A Theory of Measuring, Electing, and
Ranking, \textit{Proceedings of the National Academy of Sciences of
the USA}, vol.~104, no.~21. Pp.~8720--8725.

7. Klamler C (2006) On the Closeness Aspect of Three Voting Rules:
Borda–Copeland–Maximin, \textit{Group Decision and Negotiation},
vol.~14, issue~3. Pp.~233--240.

8. Brams SJ,  Kilgour DM,  Sanver MR (2007) A Minimax Procedure for
Electing Committees, \textit{Public Choice}, vol.~132, no.~3--4.
Pp.~401--420.

9.  Kilgour DM (2010) Approval Balloting for Multi-Winner Elections,
\textit{Handbook on Approval Voting}, Springer. Pp.~105--124.

\newpage

\begin{center}
\Large \bf Appendix
\end{center}

\textsl{ Proof of Theorem~2.}

The statements on the nonnegativity, monotonicity and superadditivity
of the functions $u, v_1$ - $v_{4}$ follow directly from their definition.
The functions $u$ and $v_1$ are not superadditive, as illustrated by Example~1.

Let us demonstrate the unanimity property. Suppose that candidate $x$
is preferable to candidate $y$ for all voters. It suffices to verify
the inequality $u(y\cup S)\leq u(x\cup S)$ for any coalition
$S\subseteq A\setminus \{x, y\}$. Denote $K=A\setminus \{x, y\}\setminus S$.
Interestingly, $h(i,x)\leq h(i,y)$ and $h(y,j)\leq h(x,j)$ for any $i, j$.
Compare the payoff matrix of the coalition $y\cup S$ against $x\cup K$

$$
\bordermatrix{
 & x & k_1 & \ldots & k_r \cr
y & 0 & h(y,k_1) & \ldots & h(y,k_r)\cr s_1 & h(s_1,x) & h(s_1,k_1)
& \ldots & h(s_1, k_r)\cr \ldots & \ldots & \ldots & \ldots &
\ldots\cr s_l & h(s_l,x) & h(s_l,k_1) & \ldots & h(s_l, k_r)}
$$
with the payoff matrix of the coalition $x\cup S$ against $y\cup K$

$$
\bordermatrix{
 & y & k_1 & \ldots & k_r\cr
x & n & h(x,k_1) & \ldots & h(x,k_r)\cr s_1 & h(s_1,y) & h(s_1,k_1)
& \ldots & h(s_1, k_r)\cr \ldots & \ldots & \ldots & \ldots &
\ldots\cr s_l & h(s_l,y) & h(s_l,k_1) & \ldots & h(s_l, k_r)}.
$$

Elements in the lower matrix are not smaller than their counterparts in the upper one.
According to Lemma~2, we have $u(y\cup S)\leq u(x\cup S)$. Hence, for candidate $x$
the Shapley value is not less than for candidate $y$. For other characteristic functions
under consideration, this property is established by analogy.

Now, we prove the monotonicity of the ranking procedure. Assume that in a ballot paper
candidate $x$ moves by one position up, whereas candidate $y$ goes by one position down.
Designate by $\widehat{u}$ and $\widehat{v_i}$ the characteristic functions resulting
from such transformation. Obviously, for any coalition $S\subseteq A\setminus \{x, y\}$
the conditions of Lemma~3 hold true. By virtue of Lemma~3, candidate $x$ (candidate $y$)
does not decrease (increase, respectively) its rank in the collective preference.

Next, check the Condorcet property. Imagine that candidate $x$ represents
the Condorcet winner and compare it with any other candidate $y$.
The following inequalities are clear for any coalition $S\subseteq A\setminus \{x, y\}$:
\begin{equation*}
u(x\cup S)>\frac{n}{2} > u(y\cup S),
\end{equation*}
\begin{equation*}
v_1(x\cup S)>\frac{n}{2} > v_1(y\cup S),
\end{equation*}
\begin{equation*}
v_i(x\cup S)> v_i(y\cup S)=0, \qquad i=2, 3.
\end{equation*}
Therefore, the Shapley value is higher for candidate $x$ than for candidate $y$.
Recall that in Example~2 the Condorcet winner does not obtain the maximum Shapley value
in the case of $v_{4}$.

To proceed, we argue the strong Condorcet property for the function $v_{2}$.
Let $w(x)\supseteq w(y)$, $l(x)\subseteq l(y)$ and $h(x,y)>n/2$.
Then $I(h(i,x)-\frac{n}{2})\leq I(h(i,y)-\frac{n}{2})$,
$I(h(y,j)-\frac{n}{2})\leq I(h(x,j)-\frac{n}{2})$ for any $i, j$.
Verify the inequality $v_2(y\cup S)\leq v_2(x\cup S)$ for any coalition
$S\subseteq A\setminus \{x, y\}$. Denote $K=A\setminus \{x, y\}\setminus S$.
Similarly to Theorem~1, compare the payoff matrix $(3)$ of the coalition $y\cup S$
against $x\cup K$ with the payoff matrix $(4)$ of the coalition $x\cup S$
against $y\cup K$.

Again, observe that elements in the lower matrix are not smaller than
their counterparts in the upper one. Thus and so, we obtain that
$v_2(y\cup S)\leq v_2(x\cup S)$. Consequently, for candidate $x$
the Shapley value is not less than for candidate $y$; this fact proves
the strong Condorcet property. Note that
$v_{2}(y \cup A\setminus \{x,y\})<v_{2}(x\cup A\setminus \{x,y\})$.
In the case of the function $v_{2}$, candidate $x$ receives a higher rank
than candidate $y$. Within the framework of Example~4, the strong Condorcet
property fails for the functions $u$, $v_1$, $v_{3}$, and $v_{4}$.

The function $v_{2}$ enjoys the strong Condorcet property, \textit{ergo}
the majority rule. Example~3 shows that this property takes no place for
$u$ and $v_{1}$. There remains one thing to do, i.e., to demonstrate
the majority rule for the functions $v_{3}$ and $v_{4}$.
Denote by $g\equiv v_2$ the function for the profile consisting of
a ballot paper with the order $a_1>a_2>...>a_m$. Consider any initial profile
of size $n$ and add $k>n$ ballot papers with the order $a_1>a_2>...>a_m$.
In this case, we naturally have
\begin{equation*}
g(S)k\leq v_i(S)\leq g(S)k + nm, \quad i=3,4,
\end{equation*}
whence it follows that
\begin{equation*}
\frac{\varphi_x(v_i)}{k}\rightarrow \varphi_x(g), \quad i=3,4,
\end{equation*}
for any candidate $x$ as $k\rightarrow\infty$.
The proof of Theorem~2 is completed.

\newpage

\begin{center}
{\small\textit{The characteristic functions $v, u, v_1$ - $v_{4}$ in Example~1.} \\
\vspace{3mm}
\begin{tabular}{|c|c|c|c|c|c|c|c|}
\hline
K  &  $v$ & $u$ & $v_1$ & $v_2$ & $v_3$ & $v_4$  \\
\hline
$\emptyset$ &  0  &  0 & 0 & 0 & 0 & 0   \\
\hline
a &  0  & 20 & 20 & 0 & 0 & 20  \\
\hline
b &  0 &  16 & 16 & 0 &  0 &  16   \\
\hline
c &  0 & 17 & 17 & 0 &  0 &  17  \\
\hline
d &  0 & 12 & 12 & 0&  0 &  12  \\
\hline
e &  0 &  21 & 21 & 0 &  0 &  21  \\
\hline
ab &  0 & 22 & 22 & 0 &  0 &  38  \\
\hline
ac &  0{.}5 &  346/15 & 20 & 0&  0 &  37  \\
\hline
ad &  0 & 20 & 20 & 0&  0 &  32  \\
\hline
ae &  0{.}5 &  23{.}5 & 21 & 0 & 0 & 41  \\
\hline
bc &  0{.}5 &  21{.}5 & 18 & 0 &  0 &  35  \\
\hline
bd &  0 & 17{.}5 & 16 &  0&  0  & 30  \\
\hline
be &  0 & 21 & 21 & 0 & 0 & 37  \\
\hline
cd &  0 & 19 & 19 & 0 & 0 & 31  \\
\hline
ce &  1 &  23 & 23 & 1 &  23 & 40  \\
\hline
de &  0{.}5 &  329/15 & 21 & 0 & 0 & 33  \\
\hline
abc &  0{.}5  &  346/15 & 22 &  0 & 0 & 57  \\
\hline
abd &  0 & 22 & 22 & 0 &  0 &  52  \\
\hline
abe &  1 & 26 & 26 &  1 &  26 &  63  \\
\hline
acd &  1 &  24 & 24 &  1 &  24 & 56  \\
\hline
ace &  1 &  27{.}5 & 27 & 1 & 27 & 64  \\
\hline
ade &  0{.}5 &   23{.}5 & 21 & 0 & 0 & 53  \\
\hline
bcd &  0{.}5 &  21{.}5 & 19 & 0 & 0 & 51 \\
\hline
bce &  1 & 25 & 25 &  2 & 48 & 65  \\
\hline
bde &  0{.}5 &  329/15 & 21 & 0 & 0 & 52  \\
\hline
cde &  1 & 23 & 23 & 1 & 23 & 54  \\
\hline
abcd &  1 & 24 & 24 &  1 & 24 & 78  \\
\hline
abce &  1 & 33 & 33 & 3 & 94 & 111  \\
\hline
abde &  1 & 28 & 28 & 2 & 54 & 91  \\
\hline
acde &  1 & 29 & 29 & 2 & 56 & 88  \\
\hline
bcde &  1 & 25 & 25 & 2 & 48 & 82  \\
\hline
abcde &  1 & 45 & 45 &  5 & 225 & 225 \\
\hline
\end{tabular}
}
\end{center}

\end{document}